\documentclass{article}
\usepackage{spconf,cite,graphicx,url,times,amsmath,amssymb,acronym,balance, hyperref}
\usepackage{graphicx, hyperref}
\usepackage{layout,cite}
\usepackage{enumitem}
\usepackage{amsmath,amssymb,amsfonts,acronym}
\usepackage{algorithmic,algorithm,balance}
\usepackage{textcomp,lipsum}
\usepackage{xcolor,color,booktabs}
\usepackage{balance, multicol, multirow, array}
\usepackage{color,soul, float}
\usepackage{url,times}
\usepackage{caption}   

\usepackage{dblfloatfix} 

\acrodef{DNN}{deep neural network}
\acrodef{RNN}{recurrent neural network}
\acrodef{CNN}{convolutional neural network}
\acrodef{FC}{fully connected}
\acrodef{CRN}{convolutional recurrent network}
\acrodef{LSTM}{long-short term memory}
\acrodef{labram}{LaBraM}
\acrodef{cbram}{CBraMod}
\acrodef{EEG}{Electroencephalography }
\acrodef{BCI}{Brain-Computer-Interface}
\acrodefplural{BCI}[BCIs]{Brain-Computer Interfaces}
\acrodef{BFMs}{Brain Foundational Models}
\acrodef{VR}{Virtual Reality}

\newcommand{\deeksha}[1]{{\color{black}#1}}

\newcommand{\un}[1]{\underline{#1}} 

\newcommand{\ub}[1]{\underline{\textbf{#1}}}
\newcommand{\revision}[1]{{\color{black}#1}}

\begin{document}

\setlength{\abovedisplayskip}{5pt}
\setlength{\belowdisplayskip}{5pt}
\setlength{\abovedisplayshortskip}{3pt}
\setlength{\belowdisplayshortskip}{3pt}


\title{Cognitive Load Estimation Using Brain Foundation Models and Interpretability for BCIs}


\name{Deeksha M. Shama$^{1*}$\thanks{$^*$ Work done during  internship at Microsoft Research.} \qquad Dimitra Emmanouilidou$^{2}$ \qquad Ivan J. Tashev$^{2}$}
\address{$^{1}$Johns Hopkins University, Baltimore, MD, USA\\ $^{2}$ Microsoft Research, Redmond, WA, USA}


\maketitle
\ninept 
\begin{abstract}
Accurately monitoring cognitive load in real time is critical for Brain-Computer Interfaces (BCIs) that adapt to user engagement and support personalized learning. Electroencephalography (EEG) offers a non-invasive, cost-effective modality for capturing neural activity, though traditional methods often struggle with cross-subject variability and task-specific preprocessing. We propose leveraging Brain Foundation Models (BFMs), large pre-trained neural networks, to extract generalizable EEG features for cognitive load estimation. We adapt BFMs for long-term EEG monitoring and show that fine-tuning a small subset of layers yields improved accuracy over the state-of-the-art. Despite their scale, BFMs allow for real‑time inference with a longer context window. To address often-overlooked interpretability challenges, we apply Partition SHAP (SHapley Additive exPlanations) to quantify feature importance. Our findings reveal consistent emphasis on prefrontal regions linked to cognitive control, while longitudinal trends suggest learning progression. These results position BFMs as efficient  and interpretable tools for continuous cognitive load monitoring in real-world BCIs.

\end{abstract}

\section{Introduction}


Cognitive load estimation plays a pivotal role in enabling intelligent systems that adapt to users’ mental states. Applications include adaptive learning systems that adjust instructional content based on cognitive state; personalized delivery platforms that respond to user engagement; and neuroadaptive games that modulate difficulty or pacing in real time. These systems show promise towards enhancing user experience, improving learning outcomes, and supporting mental well-being through passive, brain-informed adaptation. In this regard, \acp{BCI} offer a promising avenue for passive cognitive load monitoring by leveraging physiological signals such as Eye Gaze~\cite{nasrieye}, Heart Rate~\cite{chenHRV2025}, Electrocardiography (ECG) \cite{ecg}, and \ac{EEG}~\cite{beauchemin2024enhancing}. Among these, EEG stands out as a non-invasive, portable, cost-effective modality providing direct access to neural activity.

\ac{EEG} signals are inherently complex, characterized by high dimensionality, spatio-temporal dynamics, and inter-subject variability \cite{Routray2012,saha2020intra}. Traditional machine learning approaches rely on handcrafted features such as power spectral density and functional connectivity~\cite{liu2023,hassan2024eeg,tashev2024workload}, which often require extensive preprocessing and task-specific configurations\cite{kyriaki2024comprehensive}. While effective in controlled settings, these methods struggle to generalize across tasks, users, and recording conditions~\cite{gomez2021studying,kyriaki2024comprehensive}. Deep learning models, including CNNs, recurrent LSTMs, and more recently transformers, have shown promise in learning features directly from raw or minimally processed EEG~\cite{zhou2025cognitive,li2024mst,panwar2025,eeglstm}. Yet, these models are typically \textit{trained for specific tasks } from scratch, making them task-specific and thus may lack flexibility and scalability. Their performance often degrades when applied to new users or tasks, limiting their utility in real-world \acp{BCI} systems and cross-subject requirements.

In contrast, \ac{BFMs} have emerged as a new paradigm for EEG-based \acp{BCI}. These large-scale models are pre-trained on diverse EEG datasets via self-supervised objectives toward \textit{generalizable representations} of brain activity. BFMs can be adaptable to downstream tasks with minimal fine-tuning~\cite{zhang2023brant,yang2023biot,jiang2024large,wang2025cbramod,cui2024neuro}.
BFMs show great potential for short-duration EEG classification tasks, but their application to continuous cognitive load monitoring requiring long-term EEG modeling remains underexplored~\cite{tashev2024workload}. \revision{Moreover, it is crucial to ground these advanced models in  established neuroscientific cognitive theory~\cite{sweller2020cognitive}. This mandates interpretable feature analyses, often missing in large-scale  \ac{BFMs}~\cite{lai2025simple}.}

Herein, we investigate the use of BFMs for continuous cognitive‑ load monitoring and examine key challenges in scalability, generalization, interpretability. To our knowledge, this is the first work to apply BFMs to continuous cognitive‑load estimation and to analyze their behavior in a multi‑day training setting. Our contributions are:
\begin{itemize}
  \item \textbf{A scalable and cross-participant pipeline} for long-term cognitive load estimation using BFM-derived features.
  \item \textbf{A flexible group-average channel alignment}  for heterogeneous layouts, improving cross-subject generalization.
  \item \textbf{An adaptation of Partition SHAP} to interpret EEG feature and region importance, aligned with neuroscience~\cite{lundberg2017unified}.
  \item \textbf{A longitudinal analysis} across multiple days revealing learning progression w.r.t. cognitive load and other neural markers.
\end{itemize}
We find that cognitive load decreases over time while prefrontal neural relevance increases. Our results further show that BFMs, particularly LaBraM\cite{jiang2024large}, improve estimation accuracy and consistently emphasize frontal regions linked to working memory and executive function, supporting their use in real-world cognitive monitoring.

\begin{figure*}[!t]
  \centering
  \begin{minipage}[t]{0.5\textwidth}
    \centering
    \includegraphics[width=0.9\linewidth]{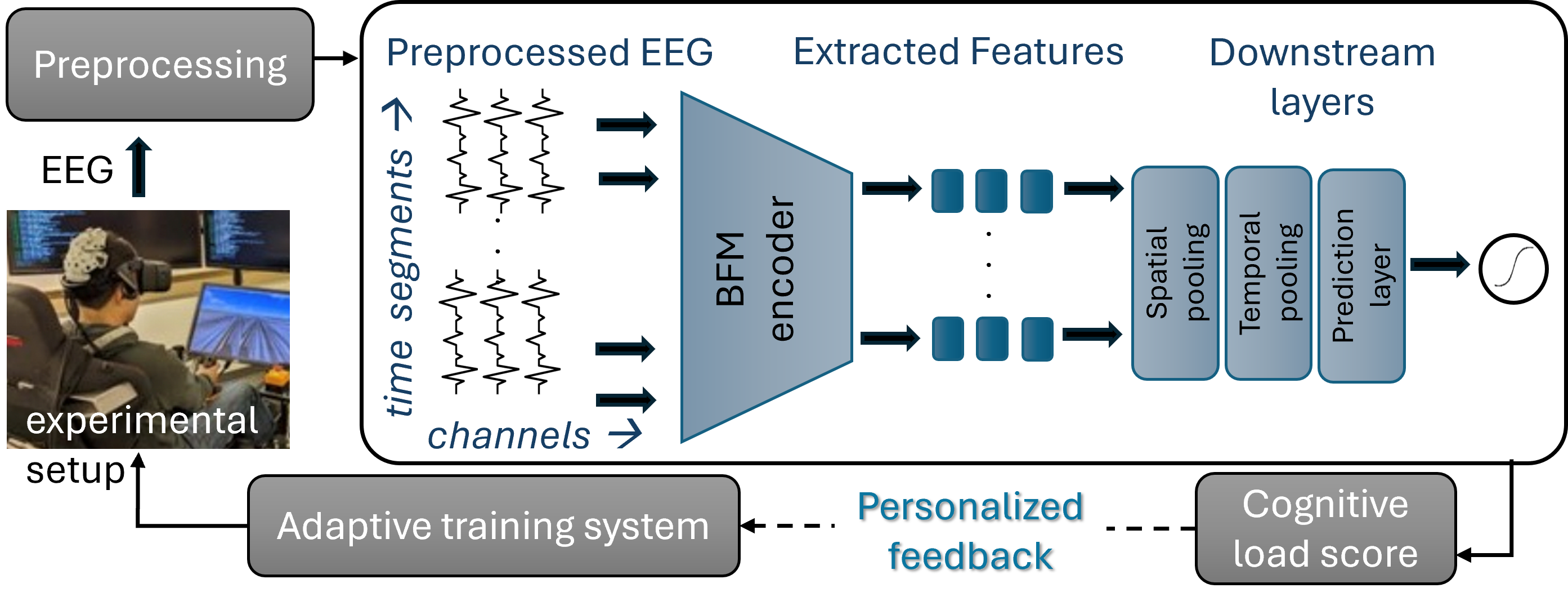}
    \captionof{figure}{Overall pipeline of cognitive load estimation with brain foundation model (BFM) in adaptive training systems.}
    \label{fig:pipeline}
  \end{minipage}\hfill
  \begin{minipage}[t]{0.48\textwidth}
    \centering
    \includegraphics[width=0.9\linewidth]{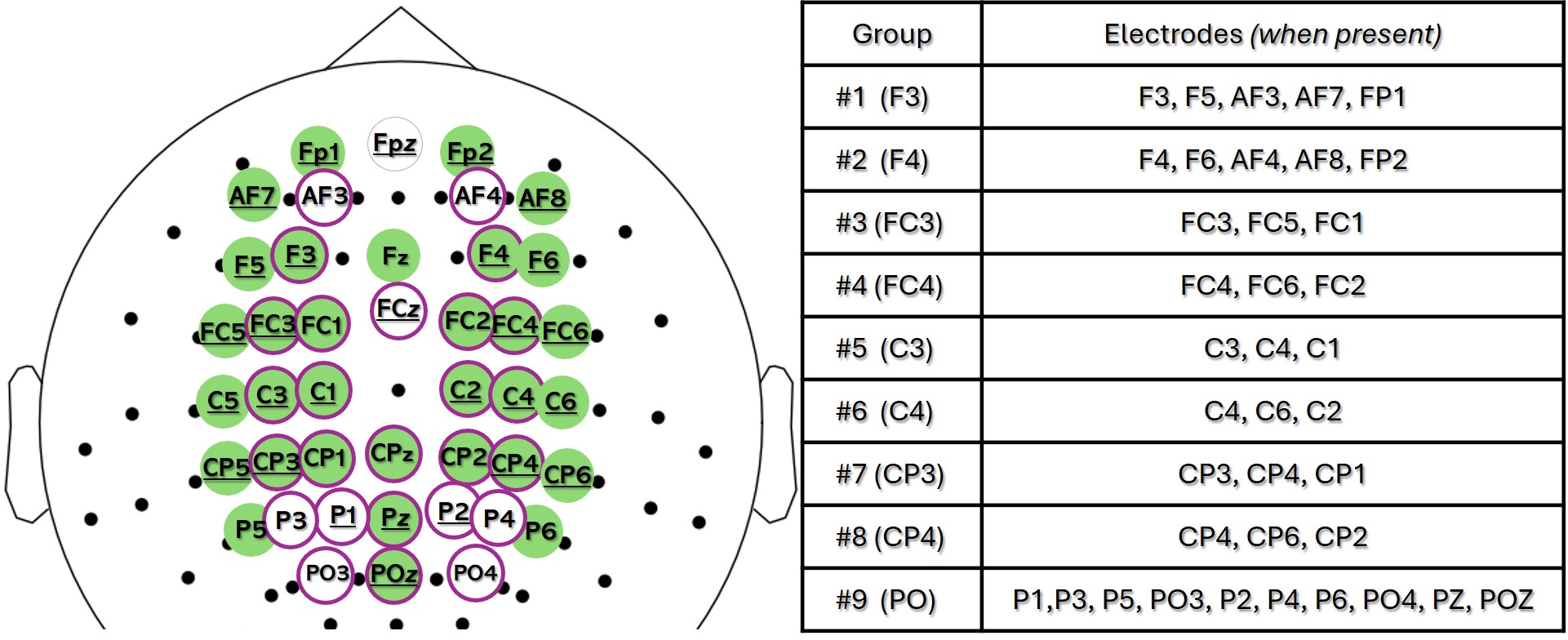}
    \captionof{figure}{Electrodes and groupings. Green fill: part of 32‑chann setup, underline: part of 28‑chann setup, purple outline: part of 26‑chann.}
    \label{fig:group}
  \end{minipage}
\vspace{-12pt}
\end{figure*}

\section{Methods}

\subsection{Data Collection and EEG Preprocessing}

Five consecutive data cohorts A, B, C, D, and E were collected over the span of 3 years, each with similar but slightly modified channel configurations or hardware. Recruited participants sat on a 6 DoF chair in a Prepar3D \ac{VR}-based flight simulator, wearing a Varjo VR3 headset with custom dry electrodes of varying \{26, 28, 32\}-channel configurations (besides  ground and reference). Custom single-pin, spring-loaded EEG electrodes were developed, gold-pinned or Ag/AgCl-coated depending on the cohort, and were embedded in an adjustable three-plate 3D printed hardware shell. The EEG sensors were connected to a BrainVision Active/Dry electrode interfaces and a LiveAmp  sampling at 500 Hz. Eye-tracking data was also collected from the integrated Varjo VR3 headset. The study was approved by the Microsoft Research Ethics IRB.

Most participants were inexperienced in VR or in flying and were presented with two tasks at varying difficulty levels (created by simulated weather conditions of wind, turbulence, visibility) \cite{tashev2024workload}: approach the runway with constant course and speed or maintain constant altitude, speed, and course on a level flight. The study included 30 participants total, who, over 5 consecutive days, completed 90 trials, each lasting $\sim$ 2 minutes (shorter if simulated aircraft crashed).  Some sessions were missed due to participant no show or discarded due to poor experimental conditions. Overall, as participants underwent consecutive training sessions, their performance was expected to improve. Participants were grouped into temporally defined cohorts, collected sequentially during the study. Cohorts D and E are considered the most stable in terms of experimental setup, reflecting improvements to the acquisition protocol, custom EEG cap design, and  signal quality management. Hence, cohort E is reserved exclusively for evaluation {(see more in Section \ref{subsection:experiments}). The cohort details:

\begin{tabular}{@{}l l l l@{}}
& Cohort A: & 2 participants,  & 26 channels \\
& Cohort B: & 6 participants,  & 26 channels \\
& Cohort C: & 6 participants,  & 28 or 32 channels (varied) \\
& Cohort D: & 11 participants, & 32 channels \\
& Cohort E: & 5 participants,  & 32 channels
\end{tabular}
\\
\\
Prior to feature extraction, EEG signals were band-pass filtered within $[0.1-75]$ Hz, with a 60 Hz notch filter to remove electrical noise, resampled to 200 Hz. On average, most trials were $\sim$100 sec long (maxed at 2 min). We extract 90 sec from the center of each trial, cropping or zero-padding to ensure uniformity (less than 10\% of segments were cropped/padded).  Segments were then split into 16-sec windows with 50\% overlap, chosen as the maximum length that could be processed by the BFMs. The resulting EEG input is denoted as follows, where number of windows $N_T$ = $(90-16)/8$+$ 1$ = $10$,  number of electrodes $N_{E}$, and $N_S$ = $16 \times 200$ = $3200$ samples:
\vspace{-4pt}
\[ x \in \mathbb{R}^{N_T \times N_{E} \times N_S}\]
\vspace{-20pt}

\subsection{Ground-truth Labels from Adaptive Training System}
\label{subsection:ATS}
In our study we make use of the comprehensive score from the Adaptive Training System (ATS) for Pilots \cite{tashev2023towards} as the \textit{objective} estimate of cognitive load, designed for pilot training in \ac{VR} flight simulators. These are \textbf{continuous ground truth label scores}, that integrate flight logs, task difficulty, participant skill, and learning rate, aligned with the cognitive load theory: when difficulty and skill are controlled, performance is largely determined by the cognitive effort exerted \cite{sweller2020cognitive}. The overall process can be seen in Fig.~\ref{fig:pipeline}.



 \renewcommand{\arraystretch}{1.1}
\begin{table*}[!th]
  \centering
\caption{Average Pearson correlation comparing BFM pipelines with spectral baselines and prior state-of-the-art deep models (left), and spatial and temporal pooling ablation  (right). Best spatial pooling per row is underlined; overall best is bolded.}
  \label{tab:results}
  \begin{minipage}{0.37\textwidth} 
    \centering
    \setlength{\tabcolsep}{6pt} 
     \vspace{-5pt}
    \captionof*{table}{(a) Feature and Estimator comparison.}
    \label{tab:baseline}
     \vspace{-3pt}
    \begin{tabular}{|c|c|c|c|c|}
      \hline
\multirow{2}{*}{\deeksha{Feature}} & \multicolumn{4}{c|}{\deeksha{Downstream Estimator layer}} \\ \cline{2-5}
                               & \deeksha{SVM} & \deeksha{Linear} & \deeksha{DNN} & \deeksha{LSTM} \\ \hline
      \deeksha{PSD} & \deeksha{0.120} & \deeksha{0.147} & \deeksha{0.133} &  \deeksha{\textbf{0.299}} \\
      \deeksha{EEGNet}                  & \deeksha{--}    & \deeksha{0.143} & \deeksha{0.157} & \deeksha{--} \\
      \deeksha{EEGConf}            & \deeksha{--}    & \deeksha{0.110} & \deeksha{0.147} & \deeksha{--} \\
      \deeksha{CBraMod}                               & \deeksha{\textbf{0.133}} & \deeksha{0.030} & \deeksha{0.108} & \deeksha{0.115} \\
      \deeksha{LaBraM}                                & \deeksha{0.113} & \deeksha{\textbf{0.281}} & \deeksha{\textbf{0.230}} & \deeksha{0.201} \\ \hline
    \end{tabular}    
  \end{minipage}%
  \hfill
  %
  \begin{minipage}{0.63\textwidth} 
    \centering
    
    \setlength{\tabcolsep}{6pt} 
    \vspace{-3pt}
     \captionof*{table}{(b) Comparison of Spatial and Temporal pooling strategies.}
    \label{tab:pooling}

    \vspace{-3pt}

    \begin{tabular}{|c|c||c|c|c||c|c|c|}
      \hline
      \multicolumn{2}{|c||}{\deeksha{Pooling dimension}} & \multicolumn{3}{c||}{\deeksha{CBraMod}} & \multicolumn{3}{c|}{\deeksha{LaBraM}} \\ \hline
      \deeksha{Temporal} & \deeksha{Spatial} & \deeksha{SVM} & \deeksha{Linear} & \deeksha{DNN}  & \deeksha{SVM} & \deeksha{Linear} & \deeksha{DNN}  \\ \hline
  \multirow{2}{*}{\deeksha{Mean}} & \deeksha{Intersec.}  & \deeksha{0.031} & \deeksha{-0.006} & \deeksha{0.058} & \deeksha{-0.052} & \un{0.094} & \deeksha{-0.006}   \\
                            & \deeksha{GroupAvg} & \un{0.137} & \un{0.005} & \un{0.070}  & \un{0.069} & \deeksha{0.078} & \ub{0.248}  \\ \hline
      \multirow{2}{*}{\deeksha{MeanStd}} & \deeksha{Intersec.} & \deeksha{0.033} & \deeksha{-0.054} & \un{0.056}  & \deeksha{-0.043} & \deeksha{0.107} & \deeksha{-0.053}  \\
                                 & \deeksha{GroupAvg} & \ub{0.145} & \un{0.015} & \un{0.056} & \un{0.072} & \ub{0.294} & \un{0.239}  \\ \hline
      \multirow{2}{*}{\deeksha{Global}} & \deeksha{Intersec.}  & \deeksha{0.010} & \ub{0.067} & \deeksha{0.101}  & \deeksha{0.007} & \deeksha{0.042} & \deeksha{-0.025} \\
                              & \deeksha{GroupAvg} & \un{0.133} & \deeksha{0.030} & \ub{0.108}  & \ub{0.113} & \un{0.281} & \un{0.230}  \\ \hline
    \end{tabular}
\vspace{-8pt}
  \end{minipage}%
\end{table*}

\subsection{Feature Extraction with Brain Foundation Models}
We extract features from preprocessed EEG using two recently released BFMs: LaBraM~\cite{jiang2024large} and CBraMod~\cite{wang2025cbramod}. 
LaBraM employs a convolutional temporal encoder, spatio-temporal trainable positional embeddings, and 12 transformers with self-attention. 
In contrast, CBraMod combines temporal and frequency-domain encoders with convolutional positional embeddings and 12 transformers with criss-cross attention.
Each encoder processes EEG at a fine resolution of one second per channel. This work adapts the encoders to create embeddings over longer, 90-sec, multi-channel EEG.
Given input $x \in \mathbb{R}^{N_T \times N_{E} \times N_S}$, each encoder outputs features $h \in \mathbb{R}^{N_T \times N_{E} \times N_d}$ per electrode channel per segment, with $N_d=200$   for both models and $N_T=10$. Multiple 1-sec outputs within 16-sec segment were compressed by averaging, matching the long-range (slower manifestation) of cognitive load during the learning task, where instantaneous fluctuations are less informative. This yields the final feature vector $h$ of size 2000$\times N_E$, with $N_E$ channels varied per montage.
%

While \ac{BFMs} may be able to handle variable electrode configurations, downstream models require fixed-size features. To achieve this, we apply spatial and temporal pooling techniques. These strategies  standardize feature dimensions across participants, enabling interpretable and computationally efficient modeling of cognitive load.

\textbf{Spatial pooling }helps reduce variability across electrode configurations. We evaluated two strategies:
(1)\textit{ Group-average}: Electrode features were averaged within 9 anatomically defined regions, shown in Fig. 2, preserving neuroscientific relevance for cognitive load estimation.
(2)\textit{  Intersection}: A single electrode per region was retained, reducing dimensionality but potentially discarding informative signals.
After spatial pooling, the features were flattened along the temporal dimension, resulting in a vector of size  $N_T \times 9 \times N_d$ = $18000$.

\textbf{Temporal pooling }was applied after spatial pooling to aggregate features over time. We evaluated three strategies:
(1) \textit{  Global pooling}: Aggregates all time steps into a single vector.
(2)\textit{  Mean pooling}: Computes the average across time.
(3) \textit{ Mean-standard deviation (MeanStd)}: Stacks mean and standard deviation over time, capturing  central tendency and variability.

\subsection{ Estimating and Explaining Cognitive Load }
\textbf{Estimation}: We estimate cognitive load as a continuous-value, similar to our ground truth, Section \ref{subsection:ATS}, with the downstream estimators:
\begin{enumerate}
\item \textbf{Linear Layer (Linear)}: with an $L1$ sparsity constraint to ensure optimal training given the large vector size. 
\item \textbf{Dense Neural Network (DNN)}: consisting of two layer neural network of sizes ($N_T \times 9 \times N_d$, $N_T\times N_d$, 1) with batch normalization and ReLU activation. 
\item \textbf{Support Vector Machine (SVM)}: with a fixed Radial Basis Function kernel (RBF) kernel. 
\end{enumerate}
%
%
%
LSTM was also used for comparison using globally pooled features, following its success in prior work \cite{tashev2024workload}, but excluded from later experiments due to incompatibility with the collapsed temporal axis.

\textbf{Explainability}: To ensure our models rely on brain-evoked features rather than other artifacts, we assess whether the frozen encoder outputs align with neuroscientific expectations. While many studies emphasize predictive accuracy, we augment our evaluation with a novel Partition SHAP-based probe to interpret model behavior~\cite{lundberg2017unified,van2022tractability}. Partition SHAP is a model-agnostic method that accounts for correlated EEG features using a domain-aware hierarchical tree to compute Owen values, an efficient approximation of Shapley values. For CBraMod and LaBraM encoders 
we perturb input EEG signals and measure changes in model output to assign relevance scores to electrodes, simulating the effect of signal loss or artifacts. These scores are aggregated per participant and model, as we evaluate alignment with expected neural patterns.

\subsection{Experiments}
\label{subsection:experiments}
We formulate cognitive load estimation as a supervised regression task, optimizing models with mean squared error (MSE) loss and evaluating performance via Pearson correlation between predicted and ground-truth scores. To assess cross-subject generalization, we used nested cross-validation (CV), where test and validation participants were never seen during training, and were drawn from the final and most stable Cohort E, as it contained five individuals who completed all 90 trials. In total, 20 CV folds were created (20 different ways of selecting one for test and one for validation, out of five).
All model feature experiments and pooling comparisons in Sections \ref{subsection:results-model}- \ref{subsection:results-pooling} used the stable Cohort D for training, with 11 participants with a 32-channel configuration.  The EEG variability experiments  and the longitudinal study in  Sections \ref{subsection:results-effect}- \ref{subsection:results-shap} incorporated all 25 participants from cohorts A-D for training. All experiments used the nested-CV from Cohort E for evaluation (testing and validation).

We benchmarked against prior work using EEG power spectral density (PSD) features across five frequency bands~\cite{tashev2024workload}, and deep learning baselines trained end-to-end, including EEGNet~\cite{eegnet} and EEGConformer~\cite{song2022eeg}, with comparable downstream layers. Unlike our modular framework, these baselines integrate spatial-temporal processing internally, limiting adaptability to classical models like SVMs and LSTMs.
All experiments were implemented in Python 3.11 with PyTorch 2.0. A tiny set of hyperparameters was considered; Linear,  with $L1$ penalties in \{0, 0.5, 1\}; SVM, with fixed RBF kernels; DNN, with Adam optimizer with a cosine annealing scheduler, and $lr$ in \{5e-4, 1e-4, 5e-5, 1e-5\}.
\section{Results}
\subsection{Model and Feature performance comparison}
\label{subsection:results-model}
We benchmarked LaBraM and CBraMod against state-of-the-art PSD features~\cite{tashev2024workload} and deep networks trained from scratch for cognitive load estimation~\cite{eegnet,song2022eeg}. Table~\ref{tab:baseline} (a) reports average cross-validated Pearson correlation over Cohorts D+E, 16 participants on a 32-electrode setup.  BFMs with group-average spatial and global temporal pooling, especially LaBraM, consistently outperform prior methods across all downstream estimator layers. LaBraM surpasses CBraMod despite the latter’s asymmetric conditional positional encoding and larger pretraining dataset, likely due to its larger encoder (6.2M vs. 5.2M params) and learned spatio-temporal positional dictionary. BFMs provide high-resolution (per-second, per-channel) encodings, enabling flexible downstream integration and improved performance with end-to-end processing lasting less than a second with all model configurations. In contrast, PSD  compresses information early, limiting performance except with LSTMs, while fixed models like EEGNet and EEGConformer require retraining per dataset and offer limited flexibility with marginal gains.
\begin{figure*}[!t]
    \centering
    \includegraphics[width=0.98\linewidth]{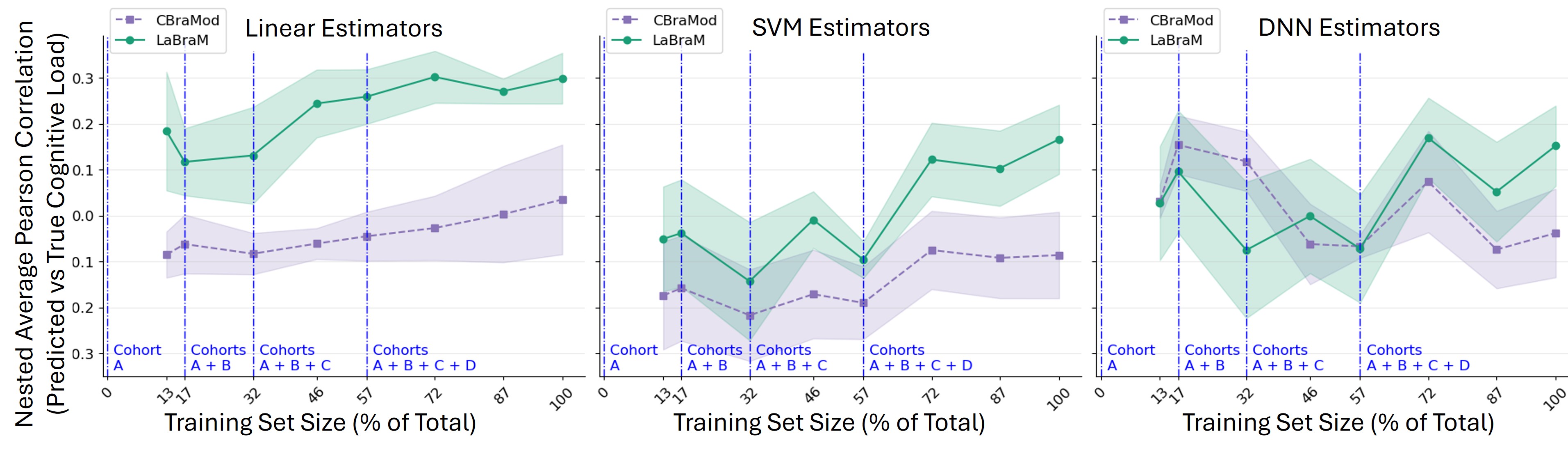}
    \vspace{-8pt}
    \caption{Average correlation trends with confidence intervals over training size (points of cohort additions shown with blue verticals). LaBraM embeddings remain robust for the cognitive load task, while a Linear layer helps avoid  overfitting to specific data cohorts. The x-axis appears not-rounded as each point reflects the addition of full participant trial sets.}
    \label{fig:data}
\end{figure*}
\begin{figure*}[!t]
    \centering
    \includegraphics[width=0.79\linewidth]{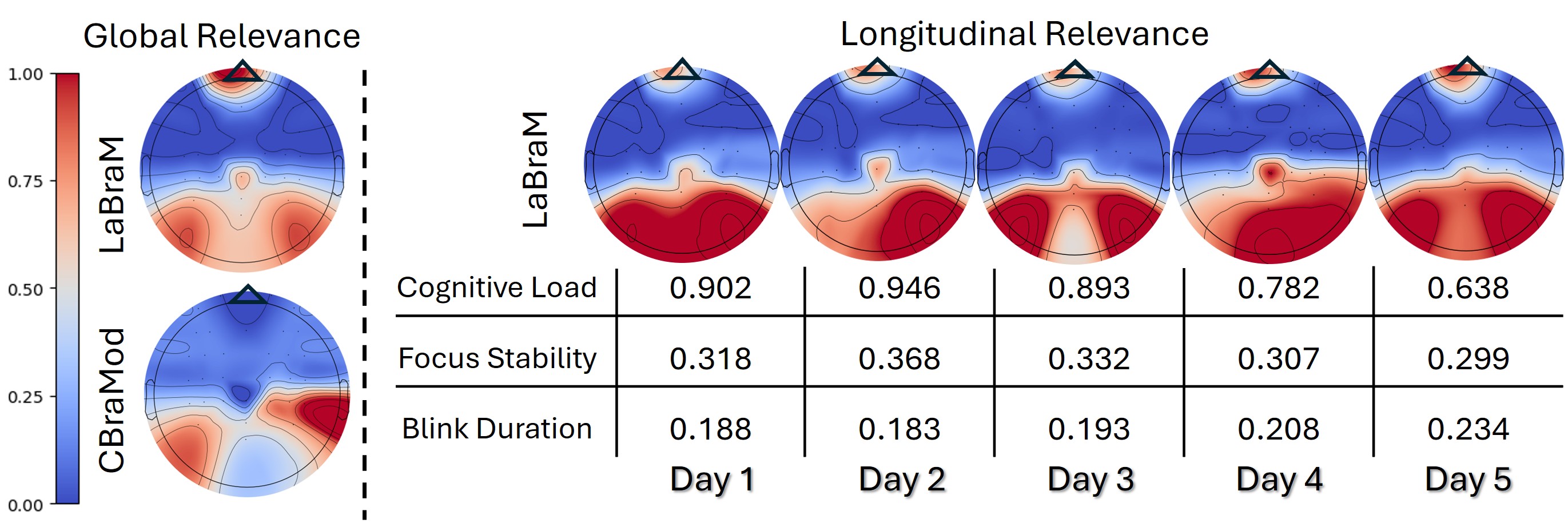}
    \vspace{-8pt}
\caption{Topomap of SHAP feature relevance, averaged globally (left) and per day (right); red indicates high importance. Daily averages of cognitive load (decreasing over time), focus stability (declining),  blink duration (increasing over time) are also shown.}
    \label{fig:shap}
\end{figure*}
\subsection{Ablation Pooling studies}
\label{subsection:results-pooling}
{We perform ablation studies to evaluate the impact of spatial and temporal pooling strategies on BFM feature aggregation across SVM, Linear, and DNN downstream estimators. LSTM was excluded, as Mean and MeanStd temporal pooling have a collapsed temporal axis.
Table~\ref{tab:pooling} (b) shows detailed results across models and pooling combinations.} No temporal pooling method consistently outperformed others: Mean and MeanStd pooling effectively reduced dimensionality, while global pooling preserved more information, offering marginally better performance. In contrast, spatial pooling had a more pronounced effect. Across all downstream layers, group-average spatial pooling consistently outperformed intersection pooling, suggesting that these BFMs capture complementary electrode-level information that intersection pooling smears. {In other words, the use of anatomically informed pooling strategies like group-average, allows for more expressive and generalizable representations,} despite heterogeneous EEG configurations.


\subsection{Effect of variable EEG configurations}
\label{subsection:results-effect}

To assess robustness to data variability, we augmented the training set of Cohort D with 14 additional participants from Cohorts A–C, which feature heterogeneous cap configurations. The same nested cross-validation setup was retained. Both CBraMod and LaBraM pipelines were retrained using the best-performing Linear layer, and average correlation scores were plotted {as a function of training set size (Fig.~\ref{fig:data}), expressed as a percentage of the full training set. Each point reflects the cumulative addition of full trial sets from participants, which results in uneven percentage increments (e.g., 17\%, 32\%, 46\%), since not all participants completed all  of their daily trials. Blue vertical lines indicating cohort additions. Confidence intervals are shown to illustrate variability across folds.} Despite increased variability, both models improved \deeksha{whereas} LaBraM achieved larger gains, indicating more robust and generalizable feature representations. The Linear layer appears robust against overfitting to cohort additions, while SVM and DNN seem more affected at the change points. Overall, 
SVM exhibited similar gaining trends across the x-axis, while DNN performance remained unaffected.

\subsection{SHAP explainability results}
\label{subsection:results-shap}
Fig.~\ref{fig:shap} (left) shows normalized global SHAP feature relevance, averaged across all days and folds, while Fig.~\ref{fig:shap} (right) presents daily SHAP maps alongside behavioral metrics (cognitive load, focus stability, blink duration). LaBraM emphasizes frontal and prefrontal regions linked to cognitive control and decision-making~\cite{euston2012role}, as well as parieto-occipital areas linked to visual working memory~\cite{koenigs2009superior,grill2004human}. In contrast, CBraMod primarily highlights parieto-occipital regions, consistent with the visual nature of the task, but lacks prefrontal emphasis, which may explain its weaker performance. These patterns were consistent across all estimators (SVM, Linear, DNN), underscoring the robustness of the extracted features and SHAP’s model-agnostic design.
{The longitudinal design, with participants returning over multiple days, enables analysis of temporal trends in cognitive load and training progression.  Fig.~\ref{fig:shap} (right) presents daily SHAP relevance maps alongside behavioral metrics of focus stability and blink duration extracted from  Varjo VR3. LaBraM shows increasing prefrontal relevance over time, while cognitive load decreases from 0.90 to 0.64 and blink duration increases from 0.19 to 0.23. 
This suggests participants became more proficient and cognitively efficient over time, while (potentially) more relaxed. These trends align with behavioral indicators such as reduced focus stability and longer blink durations, supporting the neurophysiological validity of LaBraM’s feature representations~\cite{callara2023neuronal}.} \revision{In summary,  prefrontal and parietal channels dominate  SHAP relevance, with prefrontal emphasis differences explaining LaBraM and CBraMod performance gaps.}

\section{Conclusion}
We presented a scalable pipeline for cross-subject cognitive load estimation using Brain Foundation Models (BFMs), with LaBraM outperforming CBraMod and other state-of-the-art models even with Linear estimators. Our approach generalizes across heterogeneous EEG setups and captures neurophysiologically meaningful patterns in the frontal regions. The pipeline runs in under a second on standard CPUs, supporting \revision{real‑time inference with a longer sliding window}. Our findings support the use of BFMs for cognitive load estimation, with future work aimed at integrating additional biometric signals \revision{and  complementary insights from multiple XAI methods.}

\newpage

\end{document}